\documentclass[11pt]{article}
\usepackage[top=23.4mm, bottom=23.4mm, left=23.4mm, right=23.4mm]{geometry}
\usepackage{graphicx,dcolumn,bm}
\usepackage{amssymb,amstext,amsmath}

\usepackage{graphicx}
\usepackage{dcolumn}
\usepackage{bm}
\usepackage{amssymb}
\usepackage{multirow}
\usepackage{bigstrut}
\usepackage{makecell,rotating}
\usepackage{mathrsfs}
\usepackage{booktabs}
\usepackage{threeparttable}
\usepackage{multirow}
\usepackage{subfigure}
\usepackage{colortbl}
\usepackage{epsfig}
\usepackage{threeparttable}
\usepackage{chngpage}
\usepackage{float}
\usepackage{amstext}
\usepackage{amsmath}
\usepackage{amsthm}
\usepackage{pdflscape}
\usepackage{authblk}
\usepackage{hyperref}
\usepackage{rotating}
\usepackage[graphicx]{realboxes}
\usepackage{adjustbox}
\usepackage{setspace}
\usepackage{caption}
\usepackage{xcolor}
\usepackage{xr}
\usepackage{caption}
\usepackage{lineno}
\usepackage[normalem]{ulem}

\allowdisplaybreaks
\usepackage{enumitem}
\setitemize[1]{itemsep=1pt,partopsep=0pt,parsep=\parskip,topsep=4pt}
\title{Evolutionary dynamics under coordinated reciprocity}
\author{Feipeng Zhang$^{1}$, Bingxin Lin$^{2}$, Lei Zhou$^{2*}$ and Long Wang$^{3*}$}

\date{
	$^1$ Center for Complex Systems, Xidian University, Xi’an 710071, China\\
	$^2$ School of Automation, Beijing Institute of Technology, Beijing 100081, China\\
	$^3$  Center for Systems and Control, College of Engineering, Peking University, Beijing 100871, China\\%
	$^*$ \footnotesize Corresponding authors. E-mail: leizhou@bit.edu.cn,  longwang@pku.edu.cn\\[2ex]
}

\begin{document}
	\maketitle
	
	\begin{abstract}
		Using past behaviors to guide future actions is essential for fostering cooperation in repeated social dilemmas. Traditional memory-based strategies that focus on recent interactions have yielded valuable insights into the evolution of cooperative behavior. However, as memory length increases, the complexity of analysis grows exponentially, since these strategies need to map every possible action sequence of a given length to subsequent responses. Due to their inherent reliance on exhaustive mapping and a lack of explicit information processing, it remains unclear how individuals can handle extensive interaction histories to make decisions under cognitive constraints. To fill this gap, we introduce coordinated reciprocity strategies ($CORE$), which incrementally evaluate the entire game history by tallying instances of consistent actions between individuals without storing round-to-round details. Once this consistency index surpasses a threshold, $CORE$ prescribes cooperation. Through equilibrium analysis, we derive an analytical condition under which $CORE$ constitutes an equilibrium. Moreover, our numerical results show that $CORE$ effectively promotes cooperation between variants of itself, and it outperforms a range of existing strategies including memory-$1$, memory-$2$, and those from a documented strategy library in evolutionary dynamics. Our work thus underscores the pivotal role of cumulative action consistency in enhancing cooperation, developing robust strategies, and offering cognitively low-burden information processing mechanisms in repeated social dilemmas.
	\end{abstract}
	
	\section*{Introduction}
	Social dilemmas describe situations where cooperative behavior benefits the group but comes at a personal cost, making cooperative individuals vulnerable to exploitation by free-riders \cite{axelrodbook1984evolution,Hauser:nature_unequals,McAvoy:NHB_social_goods,taylor2007transforming}. Pressing issues in human societies, such as maintaining public infrastructure, conserving natural resources, and mitigating climate change, fall into this category \cite{fehr2003nature, ostrom1990governing, orian_science_environ, pnas_Globalization_and_human_cooperation,Meng_PNAS_2025}. To address these challenges, it is essential to understand how cooperation emerges and is sustained in social dilemmas \cite{Nowak:five_rules}. Indeed, recent decades have witnessed that evolutionary game theory plays a pivotal role in uncovering the underlying mechanisms that foster cooperation \cite{ Wang:PNAS2023, anzhi_NCS_2024,anzhi_PlosCB_2023, xiaochen_nc_2023}. One of the key findings is that cooperation is more likely to evolve in situations where interactions are repeated, as individuals can observe and learn from past behaviors, adjusting their actions accordingly and eventually cultivating beneficial environment for mutual cooperation \cite{trivers:directreciprocal,  fudenberg:Folktheorem, CHENinterface,JTB_evolution_of_reciprocity,Chenxingru_Pnasnexus}.
	
	A widely adopted framework for analyzing the evolution of cooperation in repeated interactions is repeated games \cite{Boyd1987:Nopure,Boyd1989:mistakes_allows_ES}. Repeated games capture the idea that decisions are not made in isolation but unfold across multiple rounds, with the outcome in one round influencing subsequent interactions. Within this framework, one prominent line of research focuses on memory-$n$ strategies \cite{Hilbe2017:memoryn}, where individuals base their decisions on interactions of the most recent $n$ rounds. In particular, memory-$1$ strategies, which use the information of the most recent round, have been extensively studied. Strategies that effectively foster cooperation are successfully identified, such as Tit-for-Tat ($TFT$), Generous Tit-for-Tat ($GTFT$), and Win-Stay, Lose-Shift ($WSLS$) \cite{Nowak1992:TFT, Nowak1993:WSLS, Press:ZD,Prl:fairness}.
	
	In fact, the above advances largely attribute to the simplicity of memory-1 strategies, which allows an efficient search of the strategy space.  When individuals seek to make more informed decisions based on longer interaction histories, memory-$1$ strategies start failing to capture the complex patterns and subtle cues embedded in past behaviors \cite{Fischer2013:fusing_enacted_expected,HAUERT2002:JTB:Simple_adaptive_strategy,In_and_out_of_equilibrium_I,In_and_out_of_equilibrium_II,NJP_Inferring_to_cooperate,Stewart2016:smallgroups_longmamories}. One natural solution is to increase the memory length $n$, with which individuals can have access to longer interaction histories and thus make more informed decisions. However, this also introduces greater complexity \cite{Glynatsi2024,Stewart2014:Collapse_cooperation_evo}: as $n$ increases, the number of possible action sequences grows exponentially, causing a rapid expansion of the strategy space. This expansion makes the exhaustive mappings of all possible action sequences computationally prohibitive. Meanwhile, the cognitive burden of remembering long sequences of past actions is substantial, and maintaining the accuracy of such memories is often unrealistic. As a result, previous framework of memory-$n$ strategies may not adequately reflect how real-world individuals process past interactions to make decisions \cite{BOOK_Experiments_in_strategic_interaction}.
	
	Given the limitations of memory-$n$ strategies, we propose a novel approach. Instead of memorizing detailed action sequences, individuals track summaries of past interactions and use these summaries to inform their decisions. This approach offers several advantages: (i) individuals need only to store a few summary metrics rather than the entire action sequence, making it cognitively less demanding; (ii) summaries allow for the incorporation of longer interaction histories, capturing more insights without cognitively overloading; and (iii) the mappings from summaries to subsequent actions are far fewer than those required by memory-$n$ strategies, making the approach computationally feasible. Notably, the famous memory-$k$ strategy $AoN_k$ can be transformed into a strategy based on summaries \cite{Hilbe2017:memoryn,Pinheiro2014:AON}. Specifically, a single summary variable that tracks whether actions are still consistent can fully capture the logic of $AoN_k$.
	
	Here, we introduce a strategy based on coordinated reciprocity, called $CORE$, which tracks the cumulative consistency of individuals' actions throughout the entire interaction history. $CORE$ prescribes cooperation when the number of consistent actions exceeds a threshold; otherwise, it defects and resets the count of consistent actions. Unlike memory-$n$ strategies, $CORE$ summarizes past behaviors into a consistency index, allowing individuals to leverage a broader history of interactions while maintaining a manageable level of cognitive complexity. These make $CORE$ easier to implement and more computationally efficient. To evaluate the effectiveness of $CORE$, we conduct one-on-one competitions against well-established strategies. Our results show that $CORE$ not only outperforms these strategies but also more effectively promotes the evolution of cooperation than strategies with limited memories. Furthermore, when tested against machine-learning-based strategies from the Axelrod’s Python library, $CORE$ continues to show superior performance. Our results thus suggest that $CORE$ offers a powerful, low-complexity alternative to traditional memory-based strategies, and may represent a more realistic model of cooperation under the cognitive constraints faced by real-world individuals.
	
	\section*{Results}
	\paragraph{Repeated social dilemmas and $CORE$.}  We consider a social dilemma between two individuals that is repeated for an infinite number of rounds. Here, the social dilemma is described by the canonical donation game \cite{Akin2015:goodstrategies,Hilbe2017:memoryn}. In this game, individuals can choose one of the two actions, cooperation ($C$) and defection ($D$). If an individual cooperates, it incurs a cost \(c>0\) to provide a benefit \(b\) to its opponent (\(b > c\)); if an individual defects, it incurs no cost and provides no benefit. To maximize one's own payoff, the optimal choice is to defect, regardless of what the other does. However, this leads to mutual defection that yields nothing for both individuals, an outcome that is worse than if both individuals cooperated.   
	
	In repeated games, a strategy is a complete set of rules that specifies how an individual acts based on the previous outcomes. For memory-$n$ strategies, they can be represented by a tuple $(p_{h})_{h\in \mathcal{X}^n}$, where $\mathcal{X}=\{CC, CD, DC, DD\}$ is the set of outcomes in one round, $\mathcal{X}^n=\times_{i=1}^n \mathcal{X}$ is the set of histories for the previous $n$ rounds, $h$ represents one possible history, and $p_h$ denotes the probability to cooperate under history $h$. For instance, pure $TFT$ strategy is a memory-1 strategy represented by the tuple $(p_{CC}, p_{CD},p_{DC}, p_{DD})=(1, 0, 1,  0)$.
	
\begin{center}
		\begin{figure}[H]
			\centering
			\includegraphics[width=0.9\textwidth]{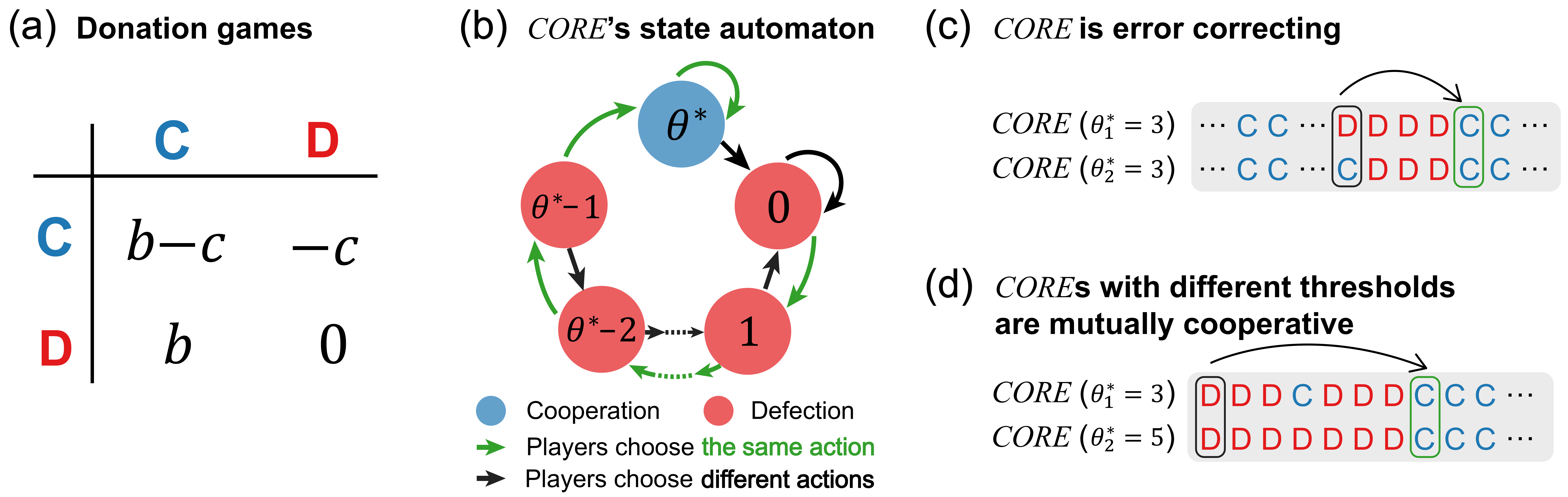}
			\caption{ \textbf{The coordinated reciprocity strategy ($CORE$) in repeated social dilemmas.} \textbf{(a),} The payoff matrix of the donation game, with player $1$'s actions listed in rows and  player $2$'s actions in columns. Cooperation ($C$) entails a cost $c$ for player to give the other player a benefit $b$ $(b>c)$, while defection ($D$) pays no cost and provides no benefit. Note that the elements in the matrix represent the payoff for player $1$. \textbf{(b),} Finite-state automata representation of $CORE$. The $CORE$ player (i.e., an individual who uses $CORE$) focuses on the cumulative consistency of actions during repeated interactions. Based on the actions of both players in each round, the $CORE$ player updates its consistency index, denoted as $\theta \in [0,\theta^*]$. When $0<\theta<\theta^*$, if both players take the same action, the index increases by one; otherwise, it decreases by one. For $\theta<\theta^*$, $CORE$ player defects. Once $\theta$ reaches the threshold $\theta^*$, $CORE$ player start cooperating. Moreover, if $CORE$ player's cooperation is exploited by the other player's defection in the previous round, $CORE$ player immediately resets $\theta$ to be zero and retaliates with defection. \textbf{(c-d),} When two $CORE$s with the same threshold meet, they can correct errors and achieve mutual cooperation. Importantly, even for $CORE$s with different thresholds, they are also mutually cooperative, contrasting with the outcome of mutual defection under two all-or-none strategies with different memory lengths. }  \label{fig1}
		\end{figure}
	\end{center}

	Unlike memory-$n$ strategies, coordinated reciprocity strategy ($CORE$) is grounded in the cumulative consistency of individuals' actions over the entire course of their interactions and needs a different representation. To this end, we introduce two parameters $\theta$ and $\theta^*$ that both take non-negative integer values. The first one $\theta\ge 0$ is called the consistency index, and the second one $\theta^*$ is the threshold and also the maximal value of $\theta$, i.e., $\theta \le \theta^*$. During repeated interactions, the consistency index $\theta$ updates itself based on the outcome of the previous round, summarizing the latest overall action consistency of individuals. Based on the latest value of $\theta$, $CORE$ then prescribes actions. 
	To explain the logic of $CORE$ in detail, we denote the current value of the consistency index as $\theta$, and denote the previous value of $\theta$ as $\theta_{\text{old}}$.
	Initially, the consistency index $\theta$ of $CORE$ is set to be zero. After each round of interaction, $\theta$ updates itself based on four cases: (i) when $0 \le \theta_{\text{old}} <\theta^*$ and individuals choose the same action in the previous round, $\theta=\theta_{\text{old}}+1$; (ii) when $\theta_{\text{old}}=\theta^*$ and individuals choose the same action, $\theta=\theta_{\text{old}}$; (iii) when $0 < \theta_{\text{old}} <\theta^*$ with individuals choosing different actions in the previous round, $\theta=\theta_{\text{old}}-1$; (iv) when $\theta_{\text{old}} = 0$ or $\theta_{\text{old}}=\theta^*$ with individuals choosing different actions, $\theta = 0$ (namely, for the former, it keeps unchanged; for the latter, it resets to zero). 
	Based on the updated value of $\theta$, $CORE$ prescribes cooperation when $\theta = \theta^*$ and defection otherwise (see Fig.~\ref{fig1}(b) for illustrations).
	
	Note that our design of $CORE$ aims to address two major challenges for the evolution of cooperation in repeated interactions: (1) exploitation by aggressive defectors, and (2) destabilization by overly permissive cooperative strategies that fail to resist the invasion of defective strategies, thereby enabling the spread of non-cooperative behavior. $CORE$ mitigates these risks by conditioning cooperation on the consistency index $\theta$, which accumulates based on the history of interactions. Cooperation is granted if and only if $\theta$ reaches a predefined threshold, ensuring that cooperative behavior is contingent on sufficiently coordinated past actions. If the threshold is not met, $CORE$ responds with defection. Such a design makes $CORE$ effectively resist aggressive strategies while also exploit unconditional or naive cooperators. Moreover, $CORE$ incorporates a strict retaliation rule: if an attempted cooperation is met with defection, the index $\theta$ is immediately reset to zero, terminating future cooperation and discouraging the opponent's opportunistic behavior. Importantly, $CORE$ is also robust to noise and implementation errors. In the case of occasional missteps, the consistency index gradually recovers through successive coordinated actions, allowing mutual cooperation to be restored within a finite number of rounds.

	Besides, in repeated games, the execution of strategies may be imperfect due to occasional mistakes. To incorporate this, we assume there exists a small error probability $\varepsilon > 0$ that individuals deviate from their intended actions.
	In other words, for most of the time (with probability $1-\varepsilon$), individuals correctly implement their intended actions ($C$ or $D$) prescribed by the strategy they use, and occasionally (with probability $\varepsilon$), they take the other action instead (e.g., defect with the intention to cooperate). 
	
	For our investigations, we focus on whether $CORE$ constitutes a successful cooperative strategy in repeated social dilemmas. To answer this, we first examine the error-correcting capabilities of $CORE$ under self-play. Then, we derive the analytical conditions under which it becomes an equilibrium. After that, we test $CORE$'s evolutionary performance when competing with other strategies in a population. For this case, two typical competing situations are considered, one with a small strategy space including at most three strategies, and the other one a large strategy space with at least seventeen strategies ($CORE$ and all deterministic memory-1 strategies). Throughout our analysis, the calculation of payoffs and cooperation rates are crucial. Except for the well-established method for games between memory-$n$ strategies, we propose a new method based on the Markov chain theory to calculate payoffs and cooperation rates of $CORE$ under self-play and against any memory-$n$ strategy (see the Supplementary Information). This method forms the foundation for our analysis of the evolutionary dynamics of $CORE$ in competition with other strategies.
	
	\paragraph{Self-play and equilibrium analysis.} To grasp a preliminary understanding of how $CORE$ behaves in repeated games, we first consider the case that two $CORE$ players interact with each other, namely, the case of self-play.  
	As noted in previous studies \cite{Fischer2013:fusing_enacted_expected,HAUERT2002:JTB:Simple_adaptive_strategy,Hilbe2015:partnersorrivals,Glynatsi2024}, the basic requirements for a cooperative strategy to succeed are to end up in a state that both players cooperate under self-play and such a state is robust to errors. To test whether $CORE$ satisfies these requirements, we calculate the average cooperation rate and payoffs when $CORE$ plays against itself. Our results show that if both players have the same threshold $\theta^*$, the average cooperation rate for each $CORE$ player is
	\begin{equation} \label{Eq.1} 
		\rho_{CORE} = \varepsilon + v_{\theta^*} (1-2\varepsilon ),
	\end{equation}
	and the associated payoff is 
	\begin{equation}\label{Eq.2}
		\pi (CORE,CORE) = (b-c)[\varepsilon + v_{\theta^*} (1-2\varepsilon )] ,        
	\end{equation} 
	where $q=\varepsilon^2+(1-\varepsilon )^2$, and $v_{\theta^*}$ is given by \\
	\begin{eqnarray}
		v_{\theta^*} &=& \left[ \frac{q^{\theta^*-1}(1-q)-(1-q)^{\theta^*}}{q^{\theta^*-1}(2q-1)} \left( \frac{1}{q} + \frac{(\theta^{*}-2)-(2\theta^*-3)q}{2q-1} + \frac{q^{\theta^*-1}}{(2q-1)(1-q)^{\theta^*-2}} + \theta^{*}-2 \right) \right. \nonumber\\
		& &~~~+ \left. \frac{1}{q} - \frac{(\theta^*-2-(2\theta^*-3)q)(1-q)}{(1-2q)^2} - \frac{q^{\theta^*-1}(1-q)}{(1-2q)^2(1-q)^{\theta^*-3}} \right]^{-1} \label{Eq.3}.
	\end{eqnarray}  
	Here, $v_{\theta^*}$ represents the probability of observing the state $\theta =\theta^*$ during repeated interactions. 
	
	In the limit of vanishing implementation errors $\varepsilon \to 0$, we can further expand \( v_{\theta^*} \) as a smooth function of \(\varepsilon\), namely, $v_{\theta^*} = 1 - C_{\theta^*} \varepsilon + \mathcal{O}(\varepsilon^2)$, where \( C_{\theta^*} > 0 \) is a coefficient that captures the influence of the first-order term of stochastic noise on the cooperative state. At this time, equation (\ref{Eq.1}) can be simplified as 
	\begin{equation}\label{Eq.4}
		\rho_{CORE} = 1 - (1 + C_{\theta^*}) \varepsilon + \mathcal{O}(\varepsilon^2),     
	\end{equation} 
	which indicates that as the probability of implementation errors approaches zero ($\varepsilon \to 0$),  $CORE$ under self-play can attain almost full cooperation.
	
\begin{center}
		\begin{figure}[H]
			\centering
			\includegraphics[width=\textwidth]{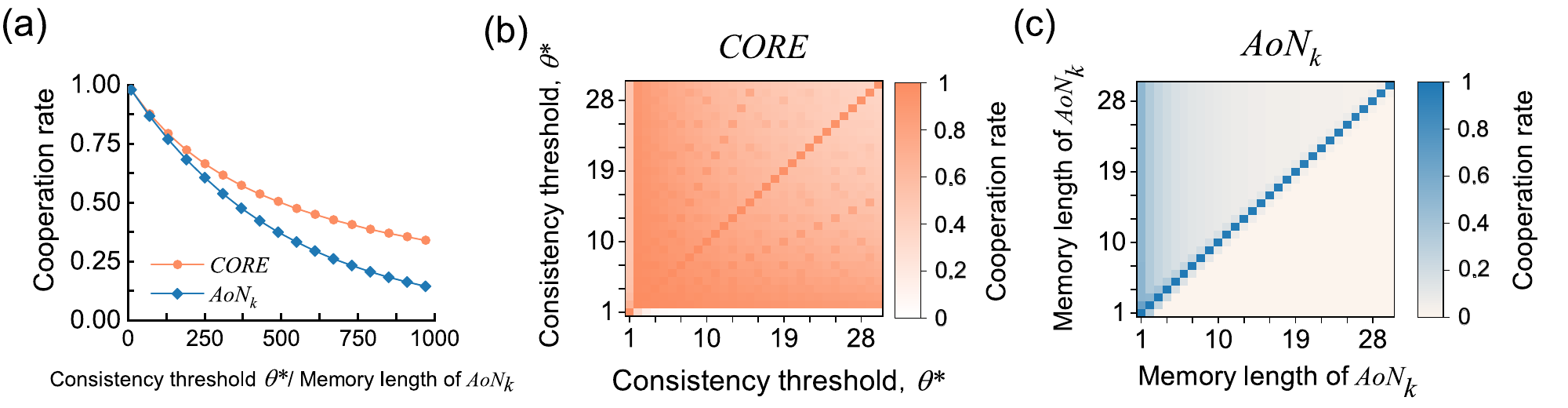}
			\caption{  \textbf{The advantages of $CORE$ over $AoN_k$ in promoting cooperation. } \textbf{(a),} The average cooperation rate under self-play shows that $CORE$ achieves a higher cooperation rate than $AoN_k$ as their corresponding key parameters varies. \textbf{(b),} For $CORE$, mutual cooperation also occurs  between $CORE$s with different $\theta^*$. Each point in the heat map represents the cooperation rate of $CORE$ corresponding to the abscissa, with the abscissa and ordinate representing the $\theta$ values of the two $CORE$s respectively. Notably, the highest cooperation rate is observed between strategies with the same threshold. \textbf{(c),} For $AoN_k$, cooperation is only observed when strategies have the same memory length. Here, we set $\varepsilon=0.1\%$.}  \label{fig2}
		\end{figure}
	\end{center}

	When the probability of implementation errors is fixed, the cooperation rate of $CORE$ under self-play depends on the error-correcting capability, which is affected by the consistency threshold $\theta^*$.
	To explore this, we calculate the cooperation rate of $CORE$ under different $\theta^*$ values. As shown in Fig.~\ref{fig2}(a), the cooperation rate of $CORE$ decreases monotonically as $\theta^*$ increases, but remains high across a broad range. To better illustrate how strong the error-correcting capability of $CORE$ is, we select one of the most successful cooperative strategies among memory-$k$ strategies in the presence of errors, namely, $AoN_k$, for comparison. Since the error-correcting capability of $AoN_k$ is closely related with the memory length $k$, we plot the cooperation rate of $AoN_k$ as a function of $k$ in Fig.~\ref{fig2}(a). 
	Our results show that when $k = \theta^*$, two $CORE$ players consistently achieve a higher cooperation rate than the $AoN_k$ players. As both $k$ and $\theta^*$ increase, the advantage of $CORE$ over $AoN_k$ becomes even more pronounced. Moreover, as shown in Fig.~\ref{fig2}(c), we find that the performance of $AoN_k$ strategies is very sensitive to the memory length $k$: when two individuals use $AoN_k$ strategies with different memory lengths, they fail to achieve mutual cooperation and the resulting cooperate rate drops drastically (even close to zero).  In contrast, two $CORE$ players with different consistency thresholds still yield high levels of cooperation for various pairs of consistency thresholds (see Fig.~\ref{fig2}(b)).
	Such an advantage highlights the great adaptability of $CORE$, which helps it sustain cooperation in heterogeneous populations. These results thus indicate that, in the presence of errors, $CORE$ outperforms $AoN_k$ in sustaining cooperation.

	Besides the capability of achieving mutual cooperation and error-correcting under self-play, successful strategies need also to be sufficiently stable when competing against other strategies. 
	Such stability can be well captured by the concept of Nash equilibrium in game theory \cite{Nash_1950}.
	A strategy is a Nash equilibrium if no single player has the incentive to deviate when others are using this strategy. Specifically, if $CORE$ is an equilibrium strategy, for any strategy $\sigma$, the payoff $\pi$ for player 1 must satisfy $\pi\left ( \sigma,CORE \right ) \le \pi\left ( CORE,CORE \right )$, meaning that the best response to a $CORE$ opponent is to use $CORE$. We show that $CORE$ is an equilibrium if
	\begin{equation}\label{SPE}
		\theta^* \ge \frac{c}{b-c}.
	\end{equation} 
	For a donation game with fixed $b$ and $c$, this inequality becomes easier to satisfy as the threshold $\theta^*$ increases. When the cost of cooperation is high, meaning the benefit-to-cost ratio ($b/c$) is low, a larger threshold is necessary for $CORE$ to become an equilibrium. Thus, increasing the consistency threshold $\theta^*$ strengthens the stability of $CORE$ in environments where cooperation is costly.
	
	\paragraph{Evolutionary analysis in a small strategy space.} 
	Becoming a Nash equilibrium ensures that no single mutant strategy can achieve a higher payoff. However, this does not imply that $CORE$ is evolutionary stable, nor does it address whether players have sufficient incentives to adopt this strategy. To better understand $CORE$'s performance, we turn to evolutionary dynamics. In evolutionary dynamics,  individuals are no longer assumed to be pure payoff-maximizers and to have unlimited cognitive abilities to figure out the optimal strategy. Instead, individuals adjust their strategies over time through learning \cite{McAvoy2022:selfishlearning, ZhouAspiration}. To gain an intuitive understanding, we consider evolutionary dynamics in a population consisting of two or three strategies, and assess whether $CORE$ has an evolutionary advantage over other strategies.
	
	We first consider a population consisting of only two strategies. Note that in both two-strategy and three-strategy scenarios, replicator dynamics provide a useful tool to describe the evolution of strategies \cite{SCHUSTER1983533:replicatordynamics}: the resulting evolutionary trajectories can be conveniently visualized using a 1-simplex or 2-simplex, in which the proportion of dominant strategies increases while the proportion of inferior strategies decreases or vanishes \cite{Schmid2022:PLCB_differentstrset}. Here, we use $1-x$ to denote the proportion of $CORE$ in the population and $x$ the proportion of the other strategy. If there exists a proportion $x^*$ such that when $x < x^*$, the population converges to $x = 0$, meaning that the population eventually becomes homogeneous and consists entirely of $CORE$, we refer to $x^*$ as the basin of attraction for $CORE$, and $1-x^*$ as the basin of attraction for the other strategy. For two strategies, a larger basin of attraction indicates that the strategy is more likely to dominate the population. Therefore, $CORE$ is considered to dominate over the other strategy if the basin of attraction of $CORE$ is larger than that of the competing strategy. 
	
\begin{figure}[H]
		\centering
		\includegraphics[width=0.8\textwidth]{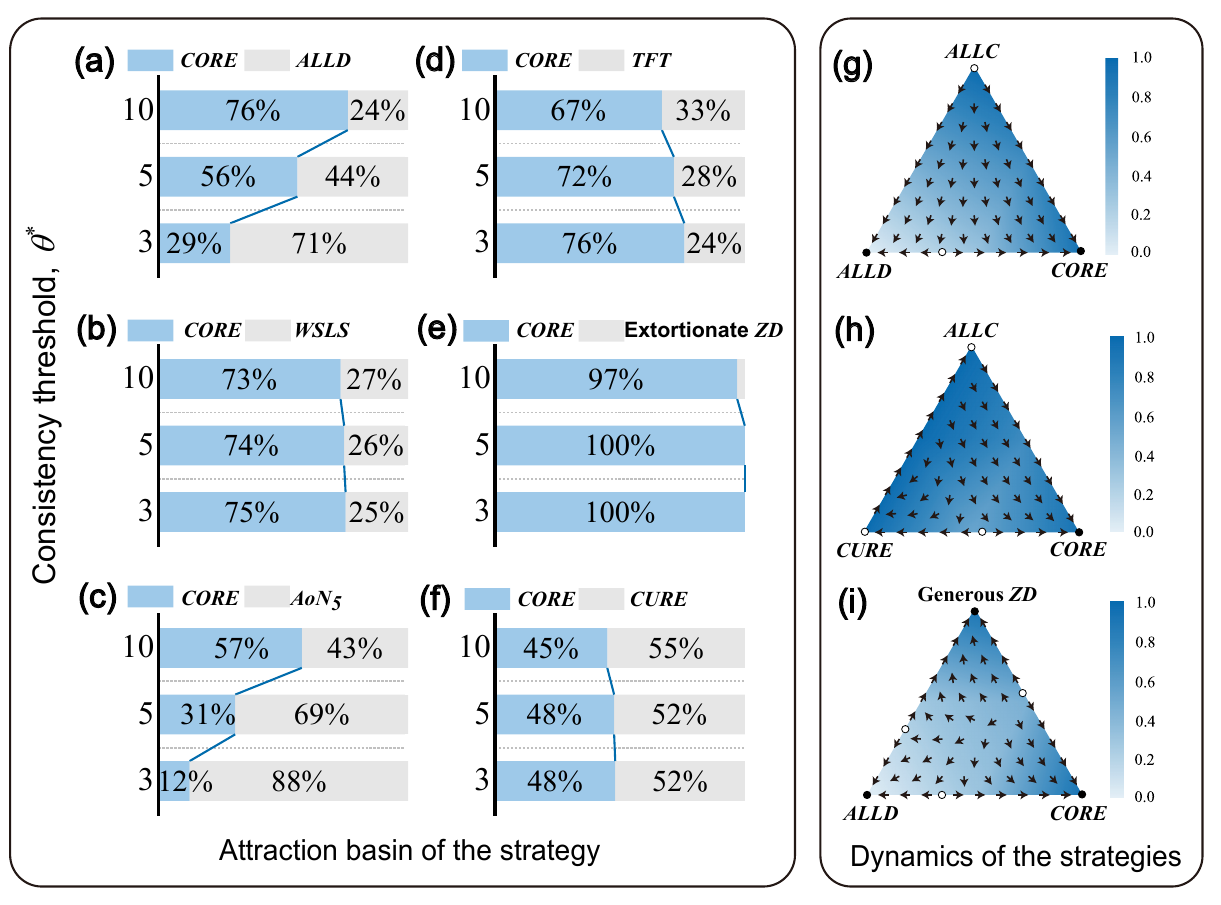}
		\caption{  \textbf{The evolutionary advantage of $CORE$ in populations with two or three strategies.} We employ replicator dynamics to model the evolution of strategies within a population. \textbf{(a)–(f)} In scenarios with two strategies, there are three potential dynamics: (i) one strategy is globally stable (dominant), (ii) each strategy is locally stable (bistable), or (iii) the two strategies coexist in a stable mixture. In the cases we consider, coexistence does not occur. When each strategy is locally stable, the basin of attraction for each strategy offers insights into its relative advantage. A strategy is considered to have a relative advantage over another if its basin of attraction is greater than $50\%$. For $TFT$ (b), $WSLS$ (d), and extortionate $ZD$ (e), $CORE$ achieves a relative advantage. For $ALLD$ and $AoN_k$, increasing $\theta^*$ can provide $CORE$ with an advantage over $ALLD$ and $AoN_k$. While $CURE$ generally shows a larger advantage than $CORE$, their advantages are nearly equal (approximately $50\%$). Parameter values are $b=1.5$ and $\varepsilon=0.1\%$. \textbf{(g)–(i)} The simplex plots, represented by triangles, illustrate the composition of the population, with each vertex corresponding to a homogeneous population of one of the three strategies. Colors denote the average payoff at a given point, while arrows indicate the direction of strategy evolution. Two stable equilibrium points exist at the vertices representing $ALLD$ and $CORE$ (h). $CORE$ is globally stable (i). Although $CURE$ may gain some advantage over $CORE$ in pairwise competition, it remains unstable against tolerant cooperative strategies like $ALLC$. After $ALLC$ dominates, the population eventually converges to a homogeneous state composed entirely of $CORE$.  All three strategies are locally stable, meaning the final convergence point depends on the initial distribution of strategies (g). Parameter values are $b=2.0$ and $\varepsilon=0.1\%$. 
		} \label{fig3}
	\end{figure}

	In detail, we evaluate the performance of $CORE$ against six widely studied strategies in Fig.~\ref{fig3}(a-f), each of which has demonstrated success in specific contexts, including $ALLD$ (always defect), $TFT$ (tit-for-tat)\cite{Nowak1992:TFT}, $WSLS$ (win-stay-lose-shift)\cite{Nowak1993:WSLS}, extortionate $ZD$ (zero-determinant strategies)\cite{Press:ZD}, $AoN_5$\cite{Hilbe2017:memoryn}, and $CURE$ (cumulative reciprocity strategy)\cite{Li2022:CURE}.
	We begin with $ALLD$, which serves as a baseline for assessing history-based strategies (see Fig.~\ref{fig3}(a)). When $\theta^* = 3$, $ALLD$ outperforms $CORE$; however, as $\theta^*$ increases, $CORE$ gains the advantage, since a higher threshold allows for more effective retaliation against defectors .
	Next, we compare $CORE$ with the history-based strategies $TFT$, $WSLS$, and the extortionate $ZD$ (see Fig.~\ref{fig3}(b, d, e)). In all cases, $CORE$ maintains a competitive advantage, though due to different reasons. Against $TFT$, the edge comes from greater robustness to implementation errors: while $TFT$ and $CORE$ earn similar payoffs when playing with each other, $TFT$ suffers heavily from errors under self-play, reducing its average payoff in the population. $CORE$ also surpasses $WSLS$ by having a larger basin of attraction, enabling it to sustain cooperation more reliably. In contrast, $ZD$ strategies deliberately restrict their opponent's payoff to extort them, but this also reduce their own payoffs, thereby hindering their population-level success when $ZD$ strategies are abundant. 
	After these, our result shows that $AoN_5$ can outperform $CORE$ when $\theta^*<5$. The reason lies in that both strategies achieve similar payoffs through mutual cooperation under self-play. But when $CORE$ plays against $AoN_5$, it is more likely to reach the cooperation threshold, making it vulnerable to the exploitations by $AoN_5$. As the cooperation threshold increases ($\theta^*>5$), $CORE$ in turn exploits $AoN_5$ and gains the upper hand (see Fig.~\ref{fig3}(c)). $CURE$ is also a strategy that evaluates the cumulative behavior of players, but relying solely on the difference between its own defections and those of its coplayers \cite{Li2022:CURE}. $CURE$ defects only when the coplayer’s number of defections exceeds its own by more than a predefined threshold; otherwise, it cooperates. This design ensures equal payoffs against any coplayer and allows $CURE$ to recover from errors quickly, enabling it to slightly outperform $CORE$ (see Fig.~\ref{fig3}(f)). However, as we discuss later, $CURE$ is vulnerable to the neutral invasion of simple cooperative strategies such as $ALLC$, which undermines its robustness in more diverse strategic environments.
	
	However, the winner in pairwise competitions is not guaranteed to retain its advantage in a multi-strategy setting, as such environments permit indirect invasions—phenomena absent in the two-strategy cases. To explore this, we examine replicator dynamics in populations consisting of three strategies (see Fig.~\ref{fig3}(g-i)). As shown in Fig. \ref{fig3}(h), while $CURE$ can outperform $CORE$, it is vulnerable to invasion by $ALLC$. As the proportion of $ALLC$ increases, $CORE$ gains a competitive edge and ultimately dominates the population. Apart from $CORE$, $ALLC$ is highly susceptible to exploitation and can be invaded by a wide range of defective strategies, which severely undermines $CURE$'s stability. We further investigate scenarios involving $ALLD$ and Generous $ZD$ (Figs. \ref{fig3}(g) and \ref{fig3}(i)), in which $CORE$ exhibits strong stability. Unlike Extortionate $ZD$, Generous $ZD$ promotes cooperation by offering benefits to its opponent. When Generous $ZD$ players are abundant in the population, this reciprocity yields higher payoffs, giving it a competitive advantage and thereby enlarging its basin of attraction.  Overall, our results show that $CORE$ can resist invasions by simple strategies and thus avoids indirect invasions by others, highlighting its stability under diverse strategic environments.

	\paragraph{Evolutionary analysis in a large strategy space.} 
	The results above indicate that $CORE$ outperforms classical strategies in both two-strategy and three-strategy settings.
	To assess the robustness of $CORE$ in a more complex competitive landscape, we expanded our analysis to a larger strategy space, focusing on evolutionary dynamics where individuals update their strategies through social imitation (see Methods). The effectiveness of this imitation process is governed by the selection strength $\beta$  \cite{Su2019:PNAS_gametransi}, with larger values enhancing the likelihood to adopt strategies that yield higher payoffs. We specifically examine scenarios with rare mutations, where the population spends most of the time in the homogeneous states with everyone adopting the same resident strategy. In this context, mutant strategies either fixate or go extinct before new mutations occur. This allows for efficient simulations, as fixation probabilities for mutants are well-defined, providing a solid framework for analyzing the evolutionary dynamics of the $CORE$.
	
	In this case, we examine whether $CORE$ is more effective at promoting cooperation than other strategies. To address this, we explore the evolutionary dynamics in populations incorporating $CORE$ and all memory-$n$ strategies up to a given complexity. To ensure a comprehensive analysis, we also extend our investigation by including $CURE$ in the population to assess whether $CORE$ can maintain its dominance. Specifically, we consider the entire strategy spaces of memory-$1$ (16 deterministic strategies) and memory-$2$ (65,536 deterministic strategies).
	
\begin{center}
		\begin{figure}[H]
			\centering
			\includegraphics[width=0.8\textwidth]{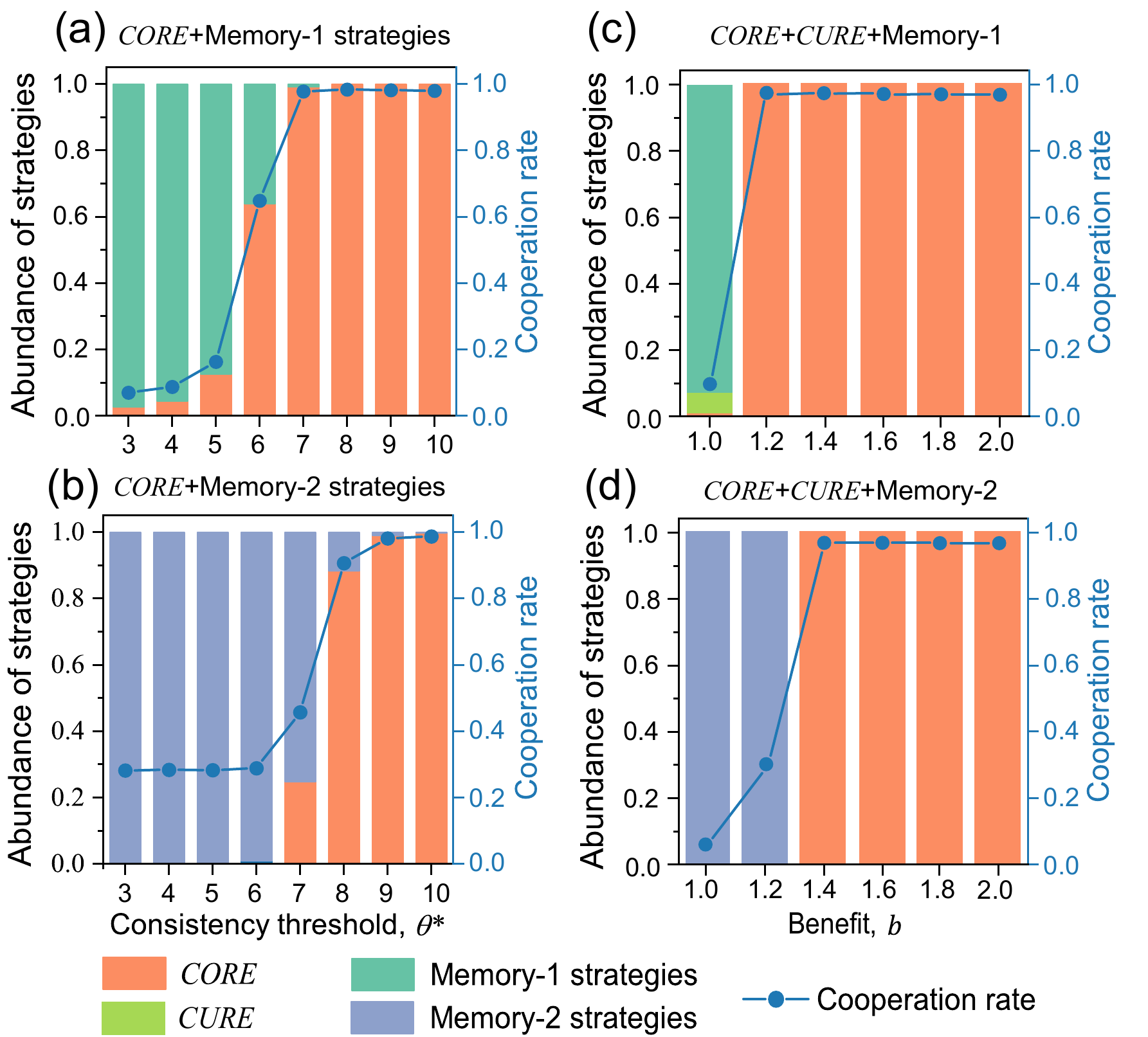}
			\caption{  \textbf{$CORE$ strategies significantly promote cooperation compared to memory-$1$ strategies, memory-$2$ strategies, and cumulative reciprocity ($CURE$) strategies.} \textbf{(a)} When considering $16$ deterministic memory-$1$ strategies and the $CORE$ strategy, as the consistency threshold ($\theta^*$) increases, $CORE$ gradually dominates the population, thereby substantially promoting cooperation. \textbf{(b)} With 65,536 deterministic Memory-2 strategies, CORE similarly drives high cooperation and dominates the population, though this requires a larger $\theta^*$.
				\textbf{(c-d)} Building on \textbf{(a)} and \textbf{(b)}, we incorporate $CURE$ strategies and observe that as the $b$ increases, $CORE$ remains the primary driver of cooperation. Parameter value: $\varepsilon=0.1\%$ and $\beta=10$. For \textbf{(a-b)}, $b=2.0$; for \textbf{(c-d)}, $\theta^*=10$ for $CORE$ and $\Delta=3$ for $CURE$.
			} \label{fig4}
		\end{figure}
	\end{center}

	As shown in Fig. \ref{fig4}(a) and (c), when memory-$1$ and memory-$2$ strategies are abundant, they fail to sustain a high level of cooperation. However, the cooperate rate increases significantly after adding $CORE$ with $\theta^*\ge 8$, highlighting the advantage of $CORE$ with a high consistency threshold over these strategies in evolutionary competing and promoting cooperation. Note that $CORE$ only accounts for less than $0.002\%$ of the total strategies in populations consisting of $CORE$ and memory-$2$ strategies. 
	This attributes to a relatively low benefit-to-cost ratio ($b/c=1.2$), where there is no equilibrium strategy in the memory-$1$ and -$2$ strategy space that can sustain cooperation. Consistent with the derived equilibrium condition ($\theta^{*}\ge c/(b-c) =5$), when the threshold exceeds a certain value, $CORE$ begins to promote a sharp increase in the cooperation rate.  
	
	Although $CURE$ can outperform $CORE$ in the two-strategy setting, $CORE$ is more likely to dominate the population when more strategies are considered (Fig. \ref{fig4}(b) and (d)). This is because $CURE$ effectively resists the invasion of selfish individuals but struggles against some unstable cooperative strategies, making it less stable.  Additionally, the increase in the cooperation rate of the population correlates with the rise in the prevalence of $CORE$, suggesting that $CORE$ directly fosters cooperation rather than indirectly facilitating it.

\begin{center}
		\begin{figure}[H]
			\centering
			\includegraphics[width=0.8\columnwidth]{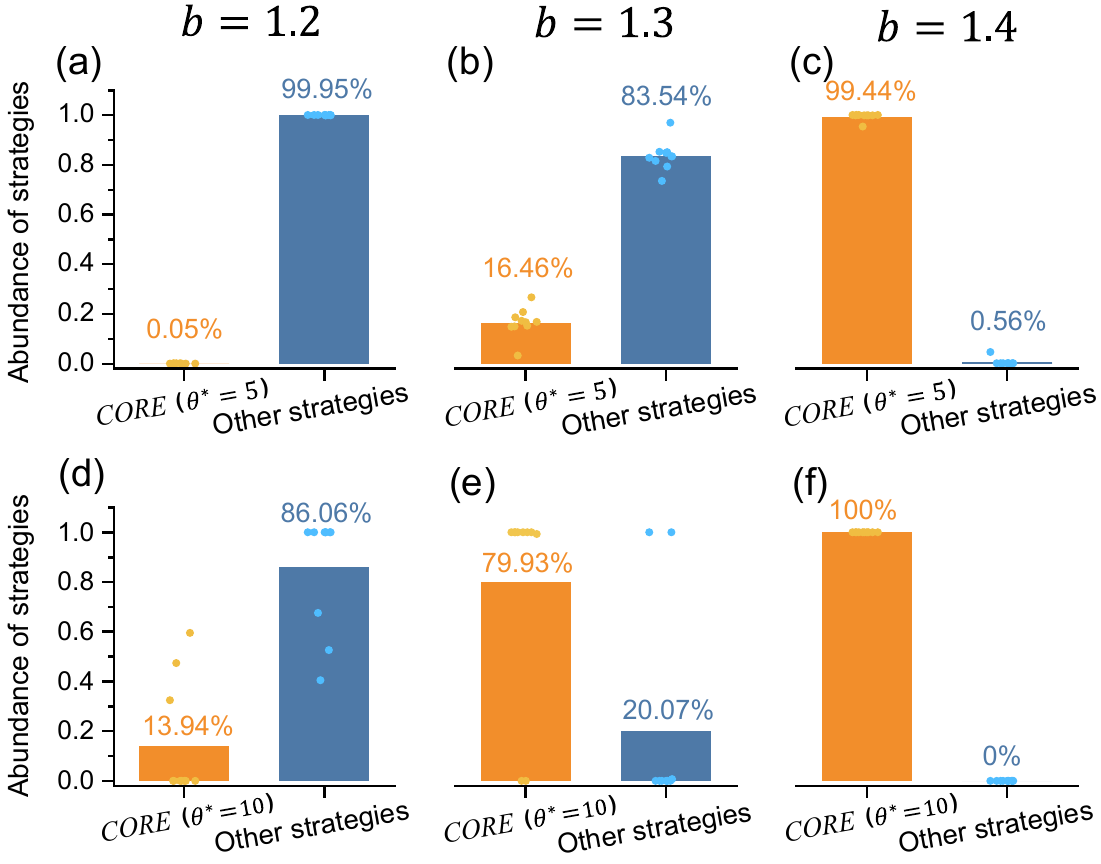}
			\caption{  \textbf{The $CORE$ strategy demonstrates a significant evolutionary advantage over the strategies included in the Axelrod Python library.}  The library includes $241$ strategies, encompassing a broad spectrum of well-established and diverse strategies that perform effectively in repeated prisoner’s dilemma games. Each bar represents the mean abundance of the $CORE$ strategy (orange) and other strategies (blue) across ten independent simulations, with dots indicating the outcomes of individual experiments.The first row (a-c) presents results for the CORE strategy with parameter \(\theta^* = 5\), while the second row (d-f) corresponds to \(\theta^* = 10\). The columns show results for different benefit-to-cost ratios (\(b\)): (a, d) \(b = 1.2\), (b, e) \(b = 1.3\), and (c, f) \(b = 1.4\). The error rate is consistently set to \(\varepsilon=0.01\).
			}  \label{fig5}
		\end{figure}
	\end{center}

	Beyond memory-based strategies, we find that $CORE$ also exhibits evolutionary advantages over a wide range of other strategies. Here, we select strategies from the Axelrod's Python library\cite{python_lib}. This library includes strategies such as $ALLD$ and $TFT$, meta-strategies like $MetaHunter$ and $MetaMajority$, as well as strategies developed by deep learning and machine learning techniques, such as $Evolved\ ANN$. To evaluate the evolutionary advantages of $CORE$, we conducted simulations where $CORE$ competed against all strategies (more than 200 strategies) from the library within a population. As shown in Fig. \ref{fig5}, our results demonstrate that $CORE$ consistently outperforms these strategies when $b\ge 1.4$. This finding suggests that $CORE$ can effectively and stably promote the evolution of cooperation within a population, while maintaining resistance to invasions by other competing strategies.

	\section*{Discussion}
	In this work, we propose a novel strategy, $CORE$, in repeated social dilemmas based on coordinated reciprocity. Unlike memory-$n$ strategies which rely on the most recent outcomes, $CORE$ evaluates the influence of decisions in all previous rounds throughout the game. Our findings show that, compared with memory-$1$ and memory-$2$ strategies, $CORE$ more effectively promotes the evolution of cooperation. 
	This results from that $CORE$ encapsulates several key features of successful cooperative strategies in repeated social dilemmas. First, it is a partner strategy that guarantees stable cooperation, ensuring that switching to any alternative strategy does not yield higher payoffs \cite{Hilbe2015:partnersorrivals}. Second, it demonstrates a remarkable capability of error handling. In scenarios involving “trembling hands” during repeated interactions, two individuals using $CORE$ with the same threshold value can restore mutual cooperation after $\theta^*$ rounds. Finally, $CORE$ proves capable of withstanding invasions by competing strategies, including both direct and indirect forms. Our work thus sheds new lights on the design of successful strategies in repeated social dilemmas beyond the framework of memory-$n$ strategies. 
	
	As mentioned above, $CORE$ summarizes all previous interaction outcomes using the consistency index, and if this index reaches a threshold, it starts to cooperate. Such a design resembles the way trust is built in human interactions: trust only develops when there has been a sufficient number of approvals. In the meanwhile, compared with memory-$n$ strategies that exhaustively map every possible outcome of the most recent $n$ rounds into an action, $CORE$ can process a lot more information in a much more efficient way without undermining its performance. Remind that the traditional approach of exhaustive mapping under memory-$n$ strategies quickly becomes infeasible to analyze as the memory length increases. This raises a new question of how to design strategies that processes information efficiently and have good performances in repeated games, which needs further explorations.  
	
	Besides, it is worth noting that increasing the consistency threshold $\theta^*$ does not necessarily lead to improved performance of $CORE$, even though higher values of $\theta^*$ tend to yield better results in the evolutionary dynamics we considered. To verify this, we let 50 different $CORE$ strategies, with $\theta^*$ values ranging from 1 to 50, to compete in a population. As shown in Fig. S3, regardless of the game parameter $b$, the most prevalent strategy is $CORE$ with a moderate value of $\theta^*$, rather than that with an extreme value (i.e., $\theta^*=1$ or $\theta^* = 50$). An intuitive explanation is that a small value of $\theta^*$ fails to effectively prevent the invasion of other strategies, while a large $\theta^*$ tends to reduce payoffs under self-play, making $CORE$ with the corresponding $\theta^*$ evolutionarily unstable. A moderate $\theta^*$ thus strikes the optimal balance, leading to the associated strategy's evolutionary success.
	
	Overall, we propose a strategy named $CORE$ that differs from traditional memory-$n$ strategies, which efficiently evaluates previous interaction outcomes by tallying instances of consistent actions between individuals without storing round-to-round details. In evolutionary competition, $CORE$ not only surpasses memory-$1$ and memory-$2$ strategies but also demonstrates advantages over strategies included in the Axelrod Python library. In the supplementary information, we also extend $CORE$ to three-player games, demonstrating its capability to facilitate mutual reciprocity in these contexts. In principle, $CORE$ can be applied to games with more than three players, provided that relevant members repeatedly participate in a shared game.
	
	\section*{Methods}
	\paragraph{Payoff Calculation in repeated games} Before calculating the payoff between $CORE$ and other strategies, we first describe the traditional payoff calculation method for memory-$n$ strategies. Specifically, a memory-$1$ strategy is represented by the vector \((p_{CC}, p_{CD}, p_{DC}, p_{DD})\), where \(p_{xy}\) denotes the probability of a player cooperating after they chose \(x \in \{C, D\}\) and their opponent chose \(y \in \{C, D\}\) in the previous round. For the four possible states \((CC, CD, DC, DD)\), the transitions between states are described by the random transition matrix:
	\[
	M(\mathbf{p}, \mathbf{q})=\left(\begin{array}{cccc}
		p^{\prime}_{CC} q^{\prime}_{CC} & p^{\prime}_{CC}(1-q^{\prime}_{CC}) & (1-p^{\prime}_{CC}) q^{\prime}_{CC} & (1-p^{\prime}_{CC})(1-q^{\prime}_{CC}) \\
		p^{\prime}_{CD} q^{\prime}_{DC} & p^{\prime}_{CD}(1-q^{\prime}_{DC}) & (1-p^{\prime}_{CD}) q^{\prime}_{DC} & (1-p^{\prime}_{CD})(1-q^{\prime}_{DC}) \\
		p^{\prime}_{DC} q^{\prime}_{CD} & p^{\prime}_{DC}(1-q^{\prime}_{CD}) & (1-p^{\prime}_{DC}) q^{\prime}_{CD} & (1-p^{\prime}_{DC})(1-q^{\prime}_{CD}) \\
		p^{\prime}_{DD} q^{\prime}_{DD} & p^{\prime}_{DD}(1-q^{\prime}_{DD}) & (1-p^{\prime}_{DD}) q^{\prime}_{DD} & (1-p^{\prime}_{DD})(1-q^{\prime}_{DD})
	\end{array}\right),
	\]
	where \(p^{\prime} = (1-\varepsilon)p + \varepsilon(1-p)\) and \(q^{\prime} = (1-\varepsilon)q + \varepsilon(1-q)\) are the effective strategies under error. This $4\times4$ matrix fully describes the dynamics of the two players after the first round. Assuming \(v(t)\) is the probability distribution of observing the states \((CC, CD, DC, DD)\) in round \(t\), and considering errors, the matrix \(M\) becomes a primitive matrix.  According to the Perron-Frobenius theorem, \(v(t)\) converges to some state \(v\) as \(t \to \infty\). This limiting probability distribution \(v\) is obtained by solving \(v = vM\), with the additional constraint that the sum of the entries of \(v\) must be 1. The expected payoffs of X and Y per round when X plays \(p\) and Y plays \(q\) are:
	
	\begin{align}
		\pi_{X} &= \langle v(p,q),(R,T,S,P) \rangle \\
		\pi_{Y} &= \langle v(p,q),(R,T,S,P) \rangle,
	\end{align}
	where \(\langle \cdot, \cdot \rangle\) denotes the standard inner product. This method can be extended to memory strategies with long but finite memory lengths, though the dimensionality of the resulting transition matrix increases. For example, fully describing the dynamics between any two memory-2 strategies requires specifying the results of the past two rounds, resulting in a \(2^4 \times 2^4\) transition matrix. For interactions between $AoN_k$ strategies, we use the method from reference 1, which is widely applicable for any \(k\) value. 
	
	$CORE$ focuses on the cumulative behavior of players, encompassing their entire history. To compute the payoffs, it is crucial to develop a mathematical description of their interaction dynamics. Instead of using an infinite number of possible past outcomes, we use consistency indices to characterize the interaction dynamics of $CORE$ players. For interactions between $CORE$ players with the same threshold \(\theta^*\), their consistency indices are always the same in the update, so a single index \(\theta\) fully describes their interactions. For interactions between $CORE$s with different thresholds \(\theta^*\), we use two possible indices to represent their current states, respectively. For interactions between $CORE$ and memory-$1$ and memory-2 strategies, we use possible indices \(\theta\) and outcomes from the past round or two rounds to describe their interactions.
	
	According to the strategy rules and the mathematical state description, their interactions form a Markov chain, and the transition matrix can fully describe their interactions. The limiting probability distribution is obtained by solving a finite-dimensional linear system of equations (for details, see supplement information). For the $CURE$s considered in this paper, we approximate their payoffs via computer simulation.
	
	The Axelrod Python library was utilized to calculate the payoffs of various strategies through simulations of multiple rounds of the Iterated Prisoner's Dilemma (IPD). Each strategy's payoff was determined based on its actions (cooperate or defect) and those of its opponents during repeated interactions. Our simulations incorporated an error rate ($\varepsilon=0.01$) to account for decision-making noise, and the final payoff for each strategy was calculated as the average over $10^4$ iterations.

	\paragraph{Evolutionary Dynamics} To model the evolutionary dynamics of strategies within a population, we use replicator dynamics when the population contains relatively few strategies. Consider an infinite population of players, each equipped with a specific strategy. Let \(x_{s_i}\) denote the proportion of players using strategy \(s_i\). Players are randomly matched in pairs to engage in repeated games, and their payoffs are calculated using the previously described method. Let \(\pi(s_i, s_j)\) represent the payoff for a player using strategy \(s_i\) when their opponent employs strategy \(s_j\). Strategies with higher fitness become more prevalent in the population, where the fitness of a strategy is proportional to the payoff it receives. Thus, the fitness of strategy \(s_i\) is
	\[
	f_{s_i} = \sum_{s_j} \pi(s_i, s_j)x_{s_j}.
	\]
	The average fitness of the population is
	\[
	\bar{f} = \sum_{s_i} f_{s_i}x_{s_i}.
	\]
	Based on these terms, replicator dynamics suggests that the proportion of each strategy changes according to the ordinary differential equation given by
	\[
	\dot{x_{s_i}} = x_{s_i}(f_{s_i} - \bar{f}).
	\]
	
	In populations with a larger number of strategies, we consider a finite population where players can update their strategies through social learning. 
	During an imitation event, two players are randomly selected from the population: a learner and a role model. Based on their strategies and the distribution of strategies in the population, we calculate the expected payoffs for the learner ($\pi _{L}$ ) and the role model ($\pi _{R}$). The learner adopts the role model's strategy with a probability given by $1/\left(1+\exp(-\beta(\pi_{R}-\pi_{L})) \right)$, where $\beta\ge 0$ is the selection strength.
	For our simulations, we model a population of size \(N\), where initially all members adopt the same strategy. In our case, the initial population consists of unconditional defectors if such a strategy exists. At each evolutionary time step, a player may switch to a new mutant strategy, randomly selected from the set of possible strategies. If the payoff of the mutant strategy is \(\pi_m(N_m)\), where \(N_m\) is the number of mutants in the population, the fixation probability of the mutant strategy can be calculated explicitly if the residents receive the payoff \(\pi_r(N_m)\):
	\[
	\Phi_{m} = \left(1 + \sum_{i=1}^{N-1} \prod_{j=1}^{i} \exp\left\{-\beta[\pi_{m}(j) - \pi_{r}(j)]\right\}\right)^{-1},
	\]
	where \(\beta \ge 0\) is the selection intensity, measuring the extent to which relative payoffs influence strategy updates. When \(\beta \to 0\), strategy updates are independent of the payoff, and the fixation probability of a mutated strategy \(\Phi_m \to 1/N\). A higher \(\beta\) value indicates that the evolutionary process more strongly favors strategies yielding higher payoffs. At each time step, according to \(\Phi_m\), the mutant either becomes the new resident strategy or goes extinct. In the next time step, another mutant strategy is introduced into the resident population. We iterate this population update process with a large number of mutation strategies, recording the current resident strategy and the resulting average cooperation rate at each step, the cooperation rate is defined as the cooperation level between the resident strategy and itself.
	\bibliography{references}   
	
	\clearpage
	
	\clearpage

	\clearpage
	\vspace*{\fill}

	\clearpage

	\clearpage	
	
	\pagebreak

\end{document}